# Translation by adaptor-helicase cycle in oligomer world

Hayato Tsuda, Osamu Narikiyo*

*Department of Physics, Kyushu University, Fukuoka 812-8581, Japan*

ABSTRACT

A mechanism of the translation in oligomer world is proposed. The translation is carried out by a minimum cycle, which is sustained by adaptors and helicases, and the first information processing in oligomer world. We expect that such a cycle actually worked in a primitive cell and can be constructed in vitro. By computer simulation we have shown that a proofreading is achieved by the fluctuation in the cell. It is rather paradoxical that the proofreading is effective for the system consisting of molecular machines with low efficiency.

* Corresponding author.
   *E-mail address:* narikiyo@phys.kyushu-u.ac.jp (O. Narikiyo)

## 1. Introduction

We have proposed an extended version (Nishio and Narikiyo, 2013) of the oligomer-world hypothesis advocated by Shimizu (Shimizu, 1996). It integrates three major hypotheses on the origin of life: RNA-world hypothesis (Atkins et al., 2011), protein-world hypothesis (Shapiro, 2007) and lipid-world hypothesis (Luisi, 2006).

Since it is a highly complicated task to model the whole cell, we focus on a part of the activities of a primitive cell in oligomer world. The first model (Nishio and Narikiyo, 2013) in this course has focused on the metabolic cycle regulated by mini-RNAs. The second (Sato and Narikiyo, 2013) has focused on the replication of mini-RNAs. This paper is the third one and focuses on the translation from mini-RNAs into mini-proteins. Throughout this paper mini-RNA and oligo-nucleotide are used in the same meaning. Similarly mini-protein and oligo-peptide are used in the same meaning.

The regulation of cell activities by mini-RNAs is significant in oligomer world. It has two aspects. One is the regulation of chemical reactions by enzymes. The other is the regulation by genetic information. The former ribozyme aspect has been discussed by the first model (Nishio and Narikiyo, 2013). In this paper we discuss the latter information aspect. More specifically we focus on the translation among various information processing in a primitive cell.

The translation in the present-day cells is carried out by highly complex molecular machines, ribosomes and t-RMAs. However, in primitive cells in oligomer world only primitive machines consisting of oligomers are available. Then we propose a minimal translation process where the functions of ribosome and t-RNA are reduced into those of helicase and adaptor as discussed in the next section. We expect that such a minimal translation process might have employed by primitive ancestor cells and will be constructed in vitro. Here we implement the translation process on computer.

The minimal model of the translation is summarized by Maynard-Smith (Maynard-Smith, 1986) and we follow it. However, his model cannot work by itself even on computer. Thus we add a molecular machine, helicase, which is necessary to sustain the translation cycle so that our revised model does work on computer at least.

## 2. Model

The necessary molecular machines for the translation are t-RNA and ribosome. To obtain a minimal model for the translation we reduce these machines into adaptor and helicase. The adaptor consists of oligo-nucleotide and the helicase consists of oligo-peptides in our model.

As in the case of the first (Nishio and Narikiyo, 2013) and second (Sato and Narikiyo, 2013) models we assume that necessary elements for our model, e.g. oligo-nucleotides, oligo-peptides and proto-ATPs, are supplied as the consequence of a molecular evolution. Moreover, as shown by the second model oligo-nucleotides become longer and longer by ligation accomplished by a special class of oligo-nucleotides. Thus it is not difficult to obtain polymers which consist of 100 monomers of nucleotides in a primitive cell.

Although the preparation of m-RMAs has to be done as another work, we assume the presence of a class of m-RNA among the polymers produced by the ligation. Such a class is selected by the information processing machines as discussed in the following. To avoid the formation of secondary structures, single-strand m-RNAs are assumed to be coated by mini-proteins (Alberts et al., 2007). Thus we can introduce a tape in which the information about the protein to be produced is written.

Our model of the translation follows that of Maynard-Smith (Maynard-Smith, 1986). However, his model does not consider the process which separates m-RNA and t-RNAs. This process is necessary and is carried out by helicases in all of the present-day cells. Thus we add a helicase, which is more primitive than the present-day helicases, to his model. Such a helicase has been also introduced into our second model (Sato and Narikiyo, 2013).

We adopt the most primitive genetic code, GNC code (Eigen et al., 1981; Maynard-Smith, 1986), where four types of codons specify four types of amino-acids, G, A, D and V.

Our adaptor is assumed to be a hairpin-structured oligo-nucleotide which is roughly one of the clover leaves of the present-day t-RNA. The adaptor has an anti-codon at the loop and carries an amino-acid at the end of the stem. The binding of the amino-acid to the adaptor is assumed to be achieved by a functional oligomer. The bridging-oligo (Tamura, 2008) is one of the candidates for such an oligomer.

To utilize the four types of adaptor whose anti-codon is CGG or CCG or CAG or CUG, we conjecture that m-RNAs have been evolutionally selected. Thus we assume that m-RNAs are the consolidations of GNC-codons in our model.

Our helicase is assumed to be a molecular motor which consists of two domains of mini-proteins and is much simpler than the present-day helicase, since some simple molecular motors consisting of mini proteins have been reported (Cordin et al., 2006;

Patel and Donmez, 2006) to have a helicase function. The fuel of the helicase, proto-ATP, is assumed to be sufficiently supplied from the environment. The helicase is assumed to bind to one of the end, G-end, of the m-RNA for simplicity. Then the separation of the m-RNA and t-RNAs begins at the G-end and finishes at the other end, C-end.

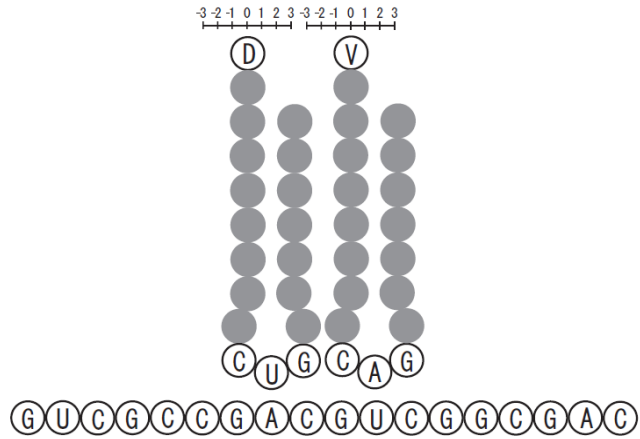

**Fig. 1.** Schematic representation of the core part of our model. The lower straight oligo-nucleotide is the m-RNA consisting of GNC codons. The upper oligo-nucleotides with hairpin structure are the adaptors. The left adaptor has CUG anti-codon and carries the amino-acid D. The right adaptor has CAG anti-codon and carries the amino-acid V. Each amino-acid can move from $-W$ to $W$ ($W = 3$) along the m-RNA. The helicase is not shown in this figure.

The implementation of our model on computer is as follows. First we prepare a single-strand m-RNA which is the consolidation of $M$ pieces of GNC-codons. Then adaptors with CNG-anti-codons try to find the appropriate codons to bind according to the Monte-Carlo rule explained later as the rule (i). During this finding process the amino-acids attached to the adaptors can form peptide-bondings according to the rule (ii) explained later. These codon-finding and bond-forming processes are repeated with the time-scales $\Delta t_{adaptor}$ and $\Delta t$. A helicase binds to the G-end of the m-RNA, after $\Delta t_{helicase}$ from the time when one adapter binds to the G-end codon, and moves along the m-RNA according to the rule (iii) explained later. After the passage of helicase from G-end of the m-RNA to C-end, we obtain mini-protein(s) translated from the m-RNA.

<u>Codon-finding rule</u> (i): Among 4 kinds of adaptors one is chosen randomly. Then the candidate codon to bind is chosen randomly from the m-RNA. This pair selection occurs at time-intervals of $\Delta t_{adaptor}$. If the chosen codon is not bound to the other adaptor, the chosen adaptor binds to the codon. If the chosen codon has been already bound to the other adaptor (A') but the amino-acid attached to A' has not form a peptide-bond yet, the chosen adaptor (A) can remove A' and bind to the codon with the probability $\exp(-\Delta E/T)$. Here $T$ represents the strength of the fluctuation in the cell. $\Delta E$ is the difference in energies, $\Delta E = E_A - E_{A'}$, where $E_A$ ($E_{A'}$) represents the energy-gain by the binding of A (A') to the candidate codon. If the center bases of the codon and the anti-codon form a Watson-Crick pair, G-C or A-U, $E_A \equiv -2$. Otherwise $E_A \equiv 2$. Thus the fluctuation plays the role of proofreading so that the correct Watson-Crick pairs are favored for long-enough waiting-time. However, once the peptide-bond is formed, the result of the translation is fixed for the codon which is bound to the amino-acid forming the bond.

Peptide-bonding rule (ii): The position of the amino-acid attached to the adaptor fluctuates. The fluctuating movement is modeled as the random walk along the m-RNA which is assumed to be a straight oligomer. The width of the random walk is given by an integer $W$ and the position is represented by an integer from $-W$ to $W$. During $\Delta t$ every amino-acid randomly moves to the adjacent position. This $\Delta t$ is the shortest time-scale in our simulation. The peptide-bonding is assumed to be formed when the distance between adjacent amino-acids becomes shortest. Here we expect that the bond is formed by the proton-shuttle mechanism (Watson et al., 2008) without the support of ribosome-like molecular machines. By forming one peptide-bonding, one amino-acid is disconnected from its adaptor. To separate the mini-protein produced by this peptide-bonding from the adaptor which still connects to the end of the mini-protein, we need a release factor. Although it is not implemented in our model, we expect that some simple oligomer can play the role of a release factor.

Helicase-motion rule (iii): We prepare three types of helicases. Type-A: The helicase motion is so quick that all the adaptors are separated from the m-RNA just after the binding of a helicase to the G-end of the m-RNA. Type-B: The characteristic time during which the helicase moves between adjacent two adaptors is $\tau$. Type-C: The characteristic time is the same as the type-B but the helicase stops at the codon not bound by an adaptor. The type-C helicase waits until an adaptor binds to the codon and restarts, while the type-B skips such a codon. Thus in the case of type-A and type-B helicases the mini-protein with full length, which has the same number of amino-acids as the number of codons of the m-RNA, may not be obtained but some fragments of the mini-protein will be obtained. On the other hand, the type-C helicase leads to the full length mini-protein.

The core part of our model is schematically shown in Fig. 1.

## 3. Simulation

We have translated the m-RNA consisting of $M$ pieces of GNC codons by our adaptor-helicase cycle. The data shown in this paper are taken for a randomly generated m-RNA with $M = 8$. Each simulation runs during $10^6 \Delta t$. The strength $T$ of the fluctuation in the cell is chosen as $T = 0.1$. All of the times are scaled by $\Delta t$.

In the following we only show the results for the type-A helicase, since the difference in the helicase type does not matter unless $\Delta t_{helicase}$ is too short.

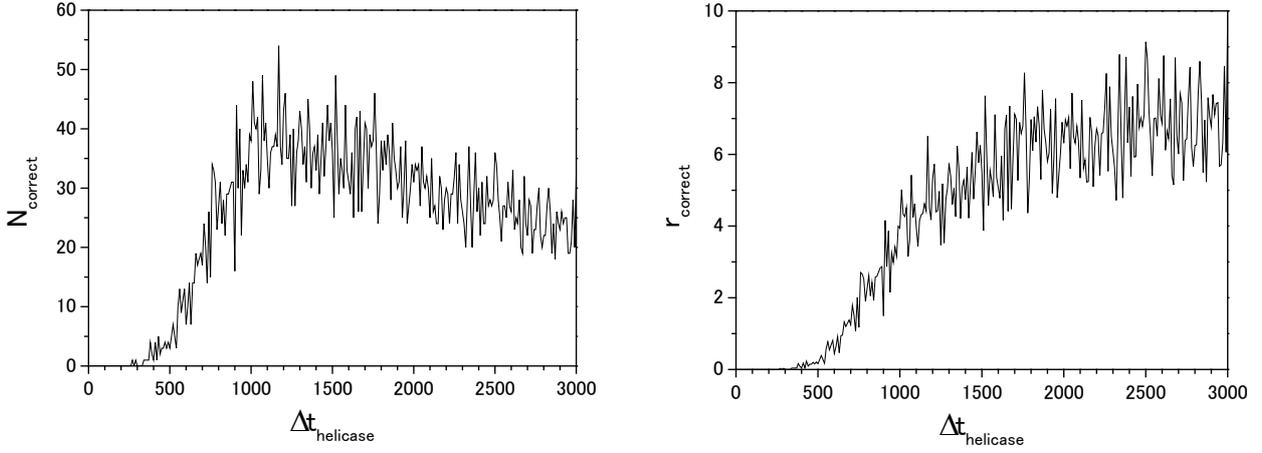

**Fig. 2.** The number $N_{correct}$ and the rate $r_{correct}$ for the correct translation when $\Delta t_{helicase}$ is changed with $\Delta t_{adaptor} = 5$ and $W = 5$.

In Fig. 2 we show the result for the correct translation when $\Delta t_{helicase}$ is changed. Here the correct translation is achieved when all of $M$ codons are translated into the amino-acids specified by the GNC code: GGC→G, GCC→A, GAC→D, GUC→V, and all of $M$ amino-acids form peptide-bonds. $N_{correct}$ is the number of the correct oligo-peptide obtained during $10^6 \Delta t$. $r_{correct}$ is the percentage of the correct oligo-peptide, $r_{correct} = 100 \times N_{correct} / N_{total}$, where $N_{total}$ is the total number of oligo-peptides obtained during $10^6 \Delta t$. The right figure shows that $r_{correct}$ increases when $\Delta t_{helicase}$ is increased. The reason is that the chances of the proofreading increase when $\Delta t_{helicase}$ is increased. On the other hand, as shown in the right figure, $N_{correct}$ decreases when $\Delta t_{helicase}$ is increased beyond $1300\Delta t$, since $\Delta t_{helicase}$ is roughly the time-interval of the events of oligo-peptide production for $\Delta t_{helicase} \gg \Delta t_{adaptor}$.

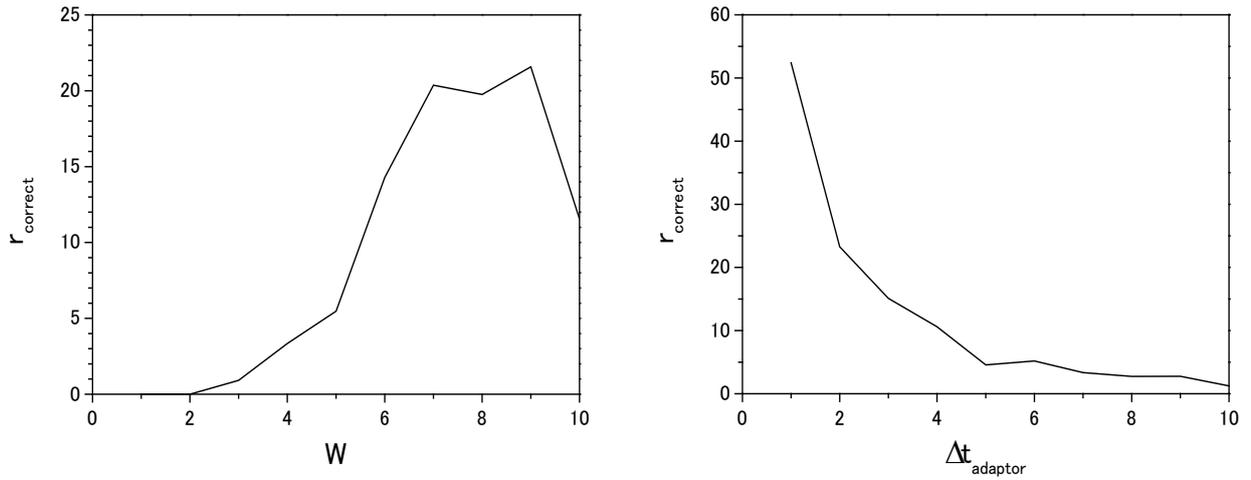

**Fig. 3.** The rate $r_{correct}$ for the correct translation when $W$ is changed with $\Delta t_{adaptor} = 5$ and $\Delta t_{helicase} = 3000$ (left) and when $\Delta t_{adaptor}$ is changed with $\Delta t_{helicase} = 3000$ and $W = 5$ (right). The results for the number $N_{correct}$ are not shown, since they are almost the same as these figures except the scales of the vertical axis.

In Fig. 3 (left) we show the result for the correct translation when $W$ is changed. $r_{correct}$ has a peak around $W = W_{optimal} \sim 8$. In the region of $W < W_{optimal}$, $r_{correct}$ decreases when $W$ is decreased. The reason is that the chances of the proofreading decreases when $W$ is decreased and the peptide-bond formation becomes easier. The proofreading is possible only before the formation of the bond. On the other hand, $r_{correct}$ decreases when $W$ is increased in the region of $W > W_{optimal}$. The reason is that the probability of forming the full-length 8-mer decreases when $W$ is increased.

Above reasoning is proven by the data shown in Table 1. As seen in Table 1-(a) for small $W = 1$, the peptide-bond is quickly formed so that the full-length 8-mer ($N_{bond} = 7$) is easily obtained. However, there is little room for the proofreading, since the bond formation is too quick. On the other hand, as seen in Table 1-(c), the formation of the full-length 8-mer becomes difficult for large $W = 10$. In-between for $W = 7$ the probability to obtain the correct 8-mer with ($N_{aa} = 8, N_{bond} = 7$), which is obtained by the correct translation through the m-RNA, becomes large as seen in Table 1-(b).

In Fig. 3 (right) we show the result for the correct translation when $\Delta t_{adaptor}$ is changed. $r_{correct}$ increases when $\Delta t_{adaptor}$ is decreased so that the chances of the proofreading increases.

As shown by the above simulations, to achieve the translation with large $r_{correct}$, the adaptor-helicase cycle with large $\Delta t_{helicase}$, small $\Delta t_{adaptor}$ and moderate $W$ is desirable. The result for such a cycle is shown in Table 1-(d) where 62.7% of the products are obtained by the correct translation through the m-RNA.

(a)

| | $N_{bond}=0$ | $N_{bond}=1$ | $N_{bond}=2$ | $N_{bond}=3$ | $N_{bond}=4$ | $N_{bond}=5$ | $N_{bond}=6$ | $N_{bond}=7$ |
|---|---|---|---|---|---|---|---|---|
| $N_{aa}=0$ | 0 | 0 | 0 | 0 | 0 | 0 | 0 | 4.6 |
| $N_{aa}=1$ | 0 | 0 | 0 | 0 | 0 | 0 | 0 | 15.5 |
| $N_{aa}=2$ | 0 | 0 | 0 | 0 | 0 | 0 | 0 | 33.4 |
| $N_{aa}=3$ | 0 | 0 | 0 | 0 | 0 | 0 | 0 | 24.9 |
| $N_{aa}=4$ | 0 | 0 | 0 | 0 | 0 | 0 | 0 | 13.7 |
| $N_{aa}=5$ | 0 | 0 | 0 | 0 | 0 | 0 | 0 | 5.5 |
| $N_{aa}=6$ | 0 | 0 | 0 | 0 | 0 | 0 | 0 | 2.1 |
| $N_{aa}=7$ | 0 | 0 | 0 | 0 | 0 | 0 | 0 | 0.3 |
| $N_{aa}=8$ | 0 | 0 | 0 | 0 | 0 | 0 | 0 | 0 |

(b)

| | $N_{bond}=0$ | $N_{bond}=1$ | $N_{bond}=2$ | $N_{bond}=3$ | $N_{bond}=4$ | $N_{bond}=5$ | $N_{bond}=6$ | $N_{bond}=7$ |
|---|---|---|---|---|---|---|---|---|
| $N_{aa}=0$ | 0 | 0 | 0 | 0 | 0 | 0 | 0 | 0 |
| $N_{aa}=1$ | 0 | 0 | 0 | 0 | 0 | 0 | 0 | 0 |
| $N_{aa}=2$ | 0 | 0 | 0 | 0 | 0 | 0 | 0 | 0.3 |
| $N_{aa}=3$ | 0 | 0 | 0 | 0 | 0 | 0 | 0 | 1.2 |
| $N_{aa}=4$ | 0 | 0 | 0 | 0 | 0 | 0 | 0.3 | 3.3 |
| $N_{aa}=5$ | 0 | 0 | 0 | 0 | 0 | 0 | 1.8 | 11.9 |
| $N_{aa}=6$ | 0 | 0 | 0 | 0 | 0 | 0 | 1.8 | 21.0 |
| $N_{aa}=7$ | 0 | 0 | 0 | 0 | 0 | 0.3 | 4.9 | 29.5 |
| $N_{aa}=8$ | 0 | 0 | 0 | 0 | 0 | 0.6 | 2.7 | 20.4 |

(c)

| | $N_{bond}=0$ | $N_{bond}=1$ | $N_{bond}=2$ | $N_{bond}=3$ | $N_{bond}=4$ | $N_{bond}=5$ | $N_{bond}=6$ | $N_{bond}=7$ |
|---|---|---|---|---|---|---|---|---|
| $N_{aa}=0$ | 0 | 0 | 0 | 0 | 0 | 0 | 0 | 0 |
| $N_{aa}=1$ | 0 | 0 | 0 | 0 | 0 | 0 | 0 | 0 |
| $N_{aa}=2$ | 0 | 0 | 0 | 0 | 0 | 0 | 0 | 0 |
| $N_{aa}=3$ | 0 | 0 | 0 | 0 | 0 | 0 | 0 | 0 |
| $N_{aa}=4$ | 0 | 0 | 0 | 0 | 0 | 0 | 0 | 0.6 |
| $N_{aa}=5$ | 0 | 0 | 0 | 0 | 0.3 | 0.3 | 1.5 | 1.5 |
| $N_{aa}=6$ | 0 | 0 | 0 | 0 | 1.2 | 3.6 | 7.6 | 4.6 |
| $N_{aa}=7$ | 0 | 0 | 0 | 0.3 | 2.7 | 7.6 | 10.6 | 8.5 |
| $N_{aa}=8$ | 0 | 0 | 0.3 | 0.9 | 4.6 | 14.3 | 17.3 | 11.6 |

(d)

| | $N_{bond}=0$ | $N_{bond}=1$ | $N_{bond}=2$ | $N_{bond}=3$ | $N_{bond}=4$ | $N_{bond}=5$ | $N_{bond}=6$ | $N_{bond}=7$ |
|---|---|---|---|---|---|---|---|---|
| $N_{aa}=0$ | 0 | 0 | 0 | 0 | 0 | 0 | 0 | 0 |
| $N_{aa}=1$ | 0 | 0 | 0 | 0 | 0 | 0 | 0 | 0 |
| $N_{aa}=2$ | 0 | 0 | 0 | 0 | 0 | 0 | 0 | 0 |
| $N_{aa}=3$ | 0 | 0 | 0 | 0 | 0 | 0 | 0 | 0 |
| $N_{aa}=4$ | 0 | 0 | 0 | 0 | 0 | 0 | 0 | 0.3 |
| $N_{aa}=5$ | 0 | 0 | 0 | 0 | 0 | 0 | 0 | 0 |
| $N_{aa}=6$ | 0 | 0 | 0 | 0 | 0 | 0 | 0.6 | 3.9 |
| $N_{aa}=7$ | 0 | 0 | 0 | 0 | 0 | 0 | 2.1 | 19.0 |
| $N_{aa}=8$ | 0 | 0 | 0 | 0 | 0 | 0.6 | 10.8 | 62.7 |

**Table 1.** The number in each cell of the table is the percentage of the product with $(N_{aa}, N_{bond})$ obtained in the simulation of Fig. 3 (left) with (a) $W=1$, (b) $W=7$ and (c) $W=10$. The data in (d) are taken with the same parameters as (b) except $\Delta t_{adaptor}=1$. Each set of product is obtained roughly by the time-interval $\Delta t_{helicase}$. $N_{aa}$ is the number of the correct correspondence between GNC-codons and amino-acids in a set of product. $N_{bond}$ is the number of the peptide-bond in a set of product. The product with $(N_{aa}=8, N_{bond}=7)$ is obtained by the correct translation through the m-RNA. A set of product with $N_{bond}<7$ may contain fragments of oligo-peptides smaller than the full-length 8-mer.

## 4. Conclusion

We have proposed a minimal model of the translation in oligomer world. The translation is sustained by the adaptor-helicase cycle. A proofreading is possible even for such a primitive model. The fluctuation in the cell is the driving force of the proofreading. The long waiting times of peptide-bond formation and helicase-binding to m-RNA are favorable for the proofreading. Paradoxically the proofreading is effective for the system consisting of molecular machines with low efficiency. Our model is expected to be the first information processing in a primitive cell and to be constructed in vitro.

**References**

Alberts, B., Johnson, A., Lewis, J., Raff, M., Roberts, K., Walter, P., 2008.
Molecular Biology of the Cell. (5-th ed.)
Garland Science, New York.

Atkins, J.F., Gesteland, R.F., Cech, T.R., 2011.
RNA Worlds, From Life's Origins to Diversity in Gene Regulation.
Cold Spring Harbor Laboratory Press, New York.

Cordin, O., Banroques, J., Tanner, N.K., Linder, P., 2006.
The DEAD-box protein family of RNA helicases.
Gene 367, 17–37.

Eigen, M., Gardiner, W., Schuster, P., Winkler-Oswatitsch, R., 1981.
The Origin of Genetic Information.
Scientific American 244 (4), 88-92.

Luisi, P.L., 2006.
The Emergence of Life, From Chemical Origins to Synthetic Biology.
Cambridge University Press, Cambridge.

Maynard-Smith, J., 1986.
The Problems of Biology.
Oxford University Press, Oxford.

Nishio, T., Narikiyo, O., 2013.
Origin and diversification of a metabolic cycle in oligomer world.
BioSystems 111, 120-126.

Patel, S.S., Donmez, I., 2006.
Mechanisms of Helicases.
J. Biological Chemistry 281, 18265-18268.

Sato, D., Narikiyo, O., 2013.
Replication of proto-RNAs sustained by ligase-helicase cycle in oligomer world.
arXiv:1306.0199.

Shapiro, R., 2007.
A Simpler Origin for Life.
Scientific American 296, (6), 46-53.

Shimizu, M., 1996.
Oligomer world: Origin of life.
Origins of Life and Evolution of Biospheres 26, 376-377.

Tamura, K., 2008.
Origin of amino acid homochirality: Relationship with the RNA world and origin of tRNA aminoacylation.
BioSystems 92, 91-98.

Watson, J.D., Baker, T.A., Bell, S.P., Gann, A., Levine, M., Losick, R., Alberts, B., 2008.
Molecular Biology of the Gene. (6-th ed.)
Cold Spring Harbor Laboratory Press, New York.